\documentclass[aip,apl,twocolumn,superscriptaddress,groupedaddress]{revtex4}  % for review and submission
\usepackage{graphicx}  % needed for figures
\usepackage{dcolumn}   % needed for some tables
\usepackage{bm}        % for math
\usepackage{amssymb}   % for math
\usepackage{epstopdf}
\setcitestyle{super}

\begin{document}

\title{High Quality Ultrathin Bi$_2$Se$_3$ Films on CaF$_2$ and CaF$_2$/Si by Molecular Beam Epitaxy with a Radio Frequency Cracker Cell}

\author{Li Zhang}
\affiliation{Department of Applied Physics, Stanford University, Stanford, CA 94305}
\affiliation{Geballe Laboratory for Advanced Materials, Stanford University, Stanford, CA 94305}
\author{Robert Hammond}
\affiliation{Geballe Laboratory for Advanced Materials, Stanford University, Stanford, CA 94305}
\author{ Merav Dolev} 
\affiliation{Department of Applied Physics, Stanford University, Stanford, CA 94305} 
\affiliation{Geballe Laboratory for Advanced Materials, Stanford University, Stanford, CA 94305}
\author{Min Liu}
\affiliation{Geballe Laboratory for Advanced Materials, Stanford University, Stanford, CA 94305}
\affiliation{Department of Physics, Stanford University, Stanford, CA 94305} 
\author{Alexander Palevski}
\affiliation{School of Physics and Astronomy, Tel Aviv University, 69978 Tel Aviv, Israel}
\author{Aharon Kapitulnik}
\affiliation{Department of Applied Physics, Stanford University, Stanford, CA 94305} 
\affiliation{Geballe Laboratory for Advanced Materials, Stanford University, Stanford, CA 94305}
\affiliation{Department of Physics, Stanford University, Stanford, CA 94305}
\date{\today}

\begin{abstract}
Here we report a method to fabricate high quality Bi$_2$Se$_3$ thin films using molecular beam epitaxy with a radio frequency cracker cell as an atomic selenium source. With Se-to-Bi ratios close to exact stoichiometry, optimal layer-by-layer growth of high quality Bi$_2$Se$_3$ thin films with smooth surfaces has been achieved on CaF$_2(111)$ substrates and Si$(111)$ substrates with a thin epitaxial CaF$_2$ buffer layer (CaF$_2$/Si). Transport measurements show a characteristic weak-antilocalization magnetoresistance in all the films, with the emergence of a weak-localization contribution in the ultrathin film limit. Quantum oscillations, attributed to the topological surface states have been observed, including in films grown on CaF$_2$/Si.

\end{abstract}

\maketitle

%While Bi$_2$Se$_3$ has been studied in the past as a narrow gap semiconductor with possible applications in thermoelectricity and infrared detectors, the discovery that this system is also a topological-insulator (TI) \cite{XLQi, HaijunZhang} has resulted in a renewed effort to fabricate it in a variety of forms, including single crystals, nanoplates, nanoribbons, and thin films. \cite{ James1, YiCui1, YiCui2,QikunXue1} 
Topological insulators (TI) are a new type of three-dimensional (3D) bulk insulating materials with surface Quantum Spin Hall Effect states protected by time reversal symmetry. In particular, Bi$_2$Se$_3$ exhibits a $0.3$ eV bulk gap and a surface state Dirac point exposed in the middle of that gap, thus identified as an ideal material-platform  for applications that use its TI nature. High quality thin films of Bi$_2$Se$_3$, with a Fermi level able to be tuned to a suitable position that is within the bulk band gap and in the surface state band simultaneously, will be key for its applications in spintronics and quantum information device physics. 

Though much effort has been devoted to the fabrication of Bi$_2$Se$_3$ thin films, primarily using molecular beam epitaxy (MBE) on a variety of substrates, \cite{QikunXue1,QikunXue2, MBE1,MBE2,MBE3,MBE4,MBE5,Ando} success has been limited due to several critical factors in the deposition process. First, selenium is known to evaporate in the form of Se$_x$ ($x\ge2$) molecules, with low sticking probability and poor chemical reactivity.\cite{selenium} As a result, Se is typically introduced during growth at a $15$:$1$ or higher ratio to Bi. However, whether it is the high volatility of Se, or a consequence of thermodynamics at $\sim 300 ^{\circ}$C substrate temperature, the resulting films always suffer from selenium vacancies and thus are n-doped in the bulk,\cite{James2,Zhanybek} leading to substantial contribution from bulk conduction, challenging the very nature of topological insulators in which only the surface state conducts. %Post growth anneal and capping the films with a thick selenium layer only partially reduces the bulk conduction. 
Therefore, making high quality Bi$_2$Se$_3$ and doped Bi$_2$Se$_3$ thin films with efficient usage of Se and desired Fermi level still remains one of the key challenges in the field. 

In this letter we report how we grew high quality Bi$_2$Se$_3$ thin films, especially ultrathin films, in a MBE system with a radio frequency (RF) cracker cell for atomic Se source. CaF$_2 (111)$ substrates were chosen for their better lattice mismatch with Bi$_2$Se$_3$ $\lesssim 7\%$, much smaller than the $13\%$ mismatch between Bi$_2$Se$_3$ and the commonly used Al$_2$O$_3 (0001)$ substrate. CaF$_2$(111) also acts as an excellent dielectric buffer layer for growth of Bi$_2$Se$_3$ films on doped-Si substrates, due to its small, $\lesssim 0.6\%$ lattice mismatch with Si(111). Using highly-doped Si $(111)$ substrates, we fabricated high quality backgated Bi$_2$Se$_3$ devices on Si substrates after an epitaxial CaF$_2$ buffer layer was first deposited (CaF$_2$/Si), which we will introduce briefly in the end. We have also grown Bi$_2$Se$_3$ films on Al$_2$O$_3(0001)$, which yielded similar results to those reported by other groups (see e.g. ref $10$ and ref $13$). Since the ultrathin Bi$_2$Se$_3$ films on CaF$_2$ and CaF$_2$/Si had superior qualities, in this paper, we concentrate on those type of films, and where appropriate, we contrast their properties with films on Al$_2$O$_3(0001)$. 

\begin{figure} [h]
\includegraphics[width=1.0 \columnwidth]{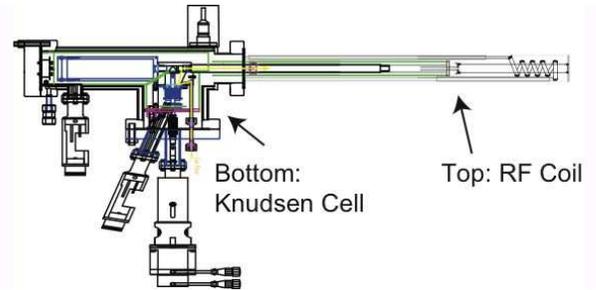}
\caption{Diagram of the RF Se cracker used in this study: on the bottom is a Knudsen cell for Se, coupled to a RF coil as the plasma section on the top.}
\label{cracker}
\end{figure}

As shown in Fig.~\ref{cracker}, a RF cracker cell for solid source materials contains two major parts that are coupled to each other through a rate valve: a regular Knudsen thermal cell (K-cell) and a RF plasma section. An inert gas plasma is first ignited in the plasma section. The K-cell is subsequently  heated up to evaporate the source material into the plasma region. The discharged source material, usually polymeric, will be dissociated into reactive atoms in the plasma section, forming a new plasma of the source material, thus enhancing the reaction with other evaporated materials and the growth of the target material. This technique has previously been applied to generate atomic phosphorus \cite{AtomicP} and arsenic\cite{AtomicAs} for doping. We adapted this technique as a composition source for uniform thin film growth. For our experiments, we installed a RF cracker cell manufactured by SVT Associates in our home-built MBE system and argon was used as the supporting gas, and this enabled us to use lower Se-to-Bi rate ratios during growth, ranging from the exact $3$:$2$ to no more than $3$:$1$, without post-growth annealing.

%CaF$_2$ is an insulator that has a face centered cubic (FCC) crystal structure with three sublattices. Its unit cell contains a simple cubic lattice formed by the F$^{-}$ ions, and the Ca$^{2+}$ ions sit in every second cube. The lattice constant is a$= 5.451 \AA$. %=
%In this paper, we will mainly describe in detail the growth and properties of Bi$_2$Se$_3$ thin films on high quality CaF$_2(111)$ substrates, and we will introduce briefly our results on growing Bi$_2$Se$_3$ films on CaF$_2$/Si, a more detailed description of the gating capabilities of the devices on CaF$_2$/Si will be given elsewhere.

Highly smooth CaF$_2 (111)$ substrates (by CrysTec GmbH) were used, with typical Atomic Force Microscopy (AFM) measured RMS-roughness $ \lesssim 0.35$ nm over a $1\times1$ $\mu$m area. Before loaded inside the MBE system, all substrates were cleaned in chloroform, acetone and methanol sequentially in an ultrasonic cleaner, $15$ minutes each. After transferred into the growth chamber, the substrates were first outgased at $600^{\circ}$C for $30$ minutes, then kept at $400^{\circ}$C and exposed to atomic oxygen for $10$ minutes for surface cleaning, with temperature of the substrates monitored in real time by a thermometer attached under the sample holder. No surface reconstructions were observed through RHEED monitoring through out this process. High purity Se $(99.999 \%)$ was evaporated from the RF cracker cell, with the rate controlled by adjusting and maintaining the opening of the rate valve accurately. High purity Bi $(99.999 \%)$ was evaporated from an electron beam source, controlled by a Quartz Crystal Microbalance (QCM), allowing us to keep both the Bi/Se ratio and the deposition rate constant throughout the growth in a vacuum of $2\times10^{-8}$ torr where a pure Se plasma (after the Ar gas was turned off) was maintained.

All Bi$_2$Se$_3$ films were grown using a two-step procedure.\cite{MBE5} In the First step, the first quintuple layer of the Bi$_2$Se$_3$ film was deposited at $197 ^\circ$C, then the sample was heated to $226 ^\circ$C at a $5 ^\circ$C/min rate, followed by the second step deposition of the film at $226  ^\circ$C. %In our study, we found $226 ^{\circ}$C to be the optimal temperature for the second step growth, which is slightly higher than the melting temperature of Se, $217 ^{\circ}$C.  
After the deposition, samples were cooled down immediately without annealing, and pure Se was deposited on some of the samples to prevent fast degradation. 
%To achieve the optimal layer-by-layer growth, a systematic study of the growth dynamics and film properties was conducted using tools both \textit{in situ}, such as real time reflection high energy electron diffraction (RHEED), and \textit{ex situ}, including X-ray diffraction (XRD), X-ray reflectivity, atomic force microscopy (AFM), scanning electron microscopy (SEM), chemical analysis and transport measurement. 
The growth procedure described above, using a $1$:$3$ Bi-to-Se ratio and a $0.35$ QL/min deposition rate, was found to yield layer-by-layer growth and highest quality ultrathin Bi$_2$Se$_3$ films on CaF$_2(111)$ and CaF$_2$/Si substrates, as will be discussed below.

\begin{figure} [h]
\includegraphics[width=1 \columnwidth]{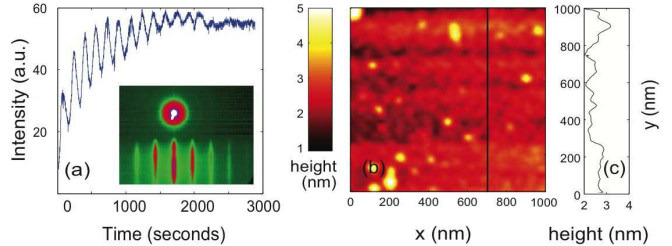}
\caption{(a) RHEED oscillations observed during growth of a $19$ QL film. Insert is the RHEED pattern of the same $19$ QL film captured immediately after deposition. (b) $1 \mu$m$\times 1 \mu$m AFM image taken on a $5$ QL thick film. (c) Depth profile of the black line that cut vertically across the sample shows that the fluctuations on top of the sample surface are smaller than $1$ nm, and the step size is $1$ nm.}
\label{afm}
\end{figure}

During the growth, the Reflection high energy electron diffraction (RHEED) patterns of the substrates faded away during the first step. From the beginning of the second step at $226^\circ$C, the peak intensity of the Bi$_2$Se$_3$ RHEED pattern would oscillate, indicating layer-by-layer growth. Fig.~\ref{afm}(a) shows that the oscillations observed during the second step growth of a $19$ quintuple layer (QL) thick film, with growth rate about $0.35$ QL/min. During the second step growth, the streaky pattern that emerged at the beginning of the second step would become more clear, indicating good crystallinity remained throughout the growth. Insert of Fig.~\ref{afm}(a) shows the RHEED pattern captured right after the deposition of the $19$ QL film shown in Fig.~\ref{afm}(a).

Topography of the Bi$_2$Se$_3$ thin films was examined by AFM using a PARK XE-70 microscope. Fig.~\ref{afm}(b) shows a typical $1 \mu$m$\times 1 \mu$m AFM scan on a $5$ QL sample. As seen from the color spread, it is a flat area with some defects and small fluctuations with amplitude smaller than $1$ nm.  This roughness is attributed to the high vapor pressure of Se, such that after the rate valve of the Se cracker cell was completely closed at the end of deposition, Se vapor in the growth chamber would still remain, until the system cooled down completely. 

\begin{figure} [h]
\includegraphics[width=1.0 \columnwidth]{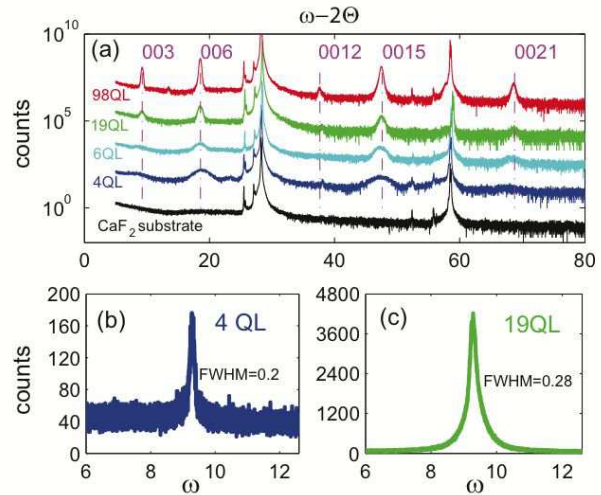}
\caption{(a) X-ray diffraction patterns (log scale) for films of different thickness. (b)X-ray rocking curve of the Bi$_2$Se$_3$ $(006)$ peak for a $4$ QL thick film (linear scale), with FWHM=$0.2^{\circ}$.  (c) X-ray rocking curve of the Bi$_2$Se$_3$ $(006)$ peak for a $19$ QL thick film, with FWHM=$0.28^{\circ}$.}
\label{xrd}
\end{figure}

Crystallinity of the Bi$_2$Se$_3$ thin films was studied using X-ray diffraction (XRD).
% performed in a PANalytical X'Pert PRO X-ray diffraction system. 
Clear c-axis orientated Bi$_2$Se$_3$ peaks were observed in films from $4$ QL up to $98$ QL thick, as shown in Fig.~\ref{xrd}(a). As shown in Fig.~\ref{xrd} (b) and (c), the FWHM of a $4$ QL film and a $19$ QL film are $0.2^{\circ}$ and $0.28^{\circ}$ respectively, indicating high crystallinity of Bi$_2$Se$_3$ was obtained from the ultrathin limit and well preserved during the layer-by-layer growth.

%For a $4$ QL thick film, the FWHM of the rocking curve for the Bi$_2$Se$_3$ $(006)$ peak, was measured to be $0.2^{\circ}$ [Fig.~\ref{xrd}(b)], indicating that high quality Bi$_2$Se$_3$ was obtained at the ultrathin limit. For a $19$ QL thick film, FWHM of the same peak, was $0.28^\circ$ [Fig.~\ref{xrd}(c)], indicating the layer-by-layer growth was stable and the high crystallinity of Bi$_2$Se$_3$ from the ultrathin limit was well preserved, consistent with the stable and streaky RHEED patterns observed during the growth, shown in Fig.~\ref{rheed}(c).

\begin{figure} [h]
\includegraphics[width=1.0 \columnwidth]{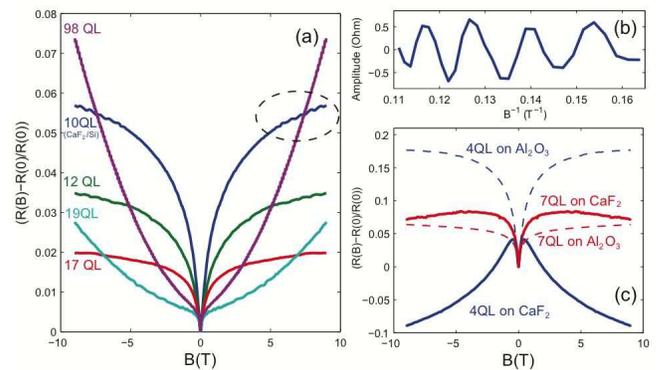}
\caption{(a) Magnetoresistance of films of different thickness on CaF$_2$, among which the $10$ QL film was on a CaF$_2$/Si$(111)$ substrate and the others were on pure CaF$_2$ substrates. The circle indicates visible quantum oscillations in the $10$ QL film, which was plotted in (b) after a smooth background was subtracted. (c) Magnetoresistance of $4$ QL and $7$ QL films on CaF$_2$ substrates (solid lines), showing a combination of both WAL and WL effects, compared to $4$ QL and $7$ QL films on Al$_2$O$_3$ $(0001)$ substrates (dashed lines) showing only WAL behavior.}
\label{mr}
\end{figure}

Transport properties of the Bi$_2$Se$_3$ thin films were measured at temperature down to $2$ K, with a magnetic field up to $9$ Tesla. The measurements were done in a $5\times5$ mm van der Pauw configuration with ohmic Ti/Au contacts. Hall effect and magnetoresistance (MR) were also measured, from which carrier density and mobility were deduced.  At $2$ K, all the films investigated, from $4$ QL up to $98$ QL, showed a strong MR cusp at low perpendicular field [Fig.~\ref{mr}(a) and (c)], which can be attributed to the characteristic weak anti-localization (WAL) behavior of the surface states in topological insulators.\cite{FuWAL,LuWAL,XueWAL} However, in the ultrathin limit, below $\sim 8$ QL, a negative MR contribution has been observed on our highest quality films on CaF$_2$ substrates. Two such examples, a $4$ QL film and a $7$ QL film, are shown in Fig.~\ref{mr} (c). Negative MR behavior has been reported before,\cite{AnisotropicMR,Bi2Se3onSi} for much thicker films ($200$ and $45$ nm/QL) in parallel fields, which was explained as a result of the locking of spin and current direction, and the field was applied in the current direction. Crossover from WAL to weak localization (WL) has also been observed in magnetically doped TI thin films, \cite{magTI} which was due to the transformation of Bi$_2$Se$_3$ from topological insulator to a topologically trivial dilute magnetic semiconductor driven by the magnetic impurities. Fundamentally different from those systems, while the emergence of negative MR (i.e. a WL contribution) in our ultrathin films in perpendicular fields is still not fully understood, it has been proposed via several models that it originates from the bulk bands. For example, quantization of the bulk states at ultrathin limit was predicted to exhibit WL if the Fermi level is close to the bottom of the bulk conduction band, and is crossing a small but finite number of these quantized bands.\cite{HaiZhouLu,Garate}  Another possibility is when the thin film is of high quality and the difference between the two opposite topological surfaces is apparent, as to give rise to a gapped Dirac hyperbolas with Rashba-like splittings in energy spectrum due to coupling between opposite topological surfaces. \cite{Wen-Yu,Eremeev}  More detailed studies of the MR, and the relevance of each of these models to our data will be discussed in a forthcoming publication. For now, it is sufficient to note that any of these explanations requires high quality surfaces, at the interface both with the substrate and with vacuum, which is the subject of the present paper. %Furthermore, quantum oscillations that originate from the surface state, have been observed in films as thin as $4$ QL on CaF$_2$ substrates, again, a signature of high quality films. 

\begin{figure} [h]
\includegraphics[width=1.0 \columnwidth]{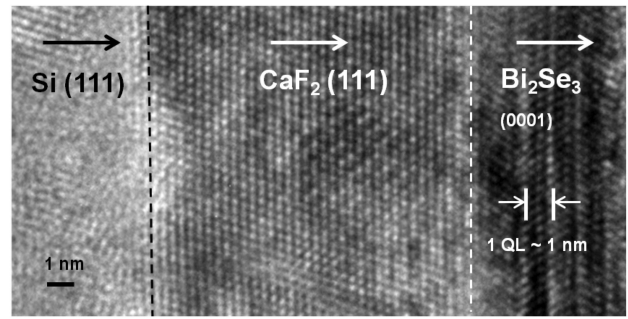}
\caption{Cross-sectional TEM image of a $5$ QL thick Bi$_2$Se$_3$ film grown on Si with a CaF$_2$ buffer layer about $12$ nm thick.}
\label{tem}
\end{figure}

%After establishing the procedure of growing high quality Bi$_2$Se$_3$ thin films on CaF$_2$ substrates, 
We also fabricated Bi$_2$Se$_3$ thin films on p-doped Si$(111)$ substrates with CaF$_2$ as buffer layers (CaF$_2$/Si). The Si$(111)$ substrates were cleaned in buffered oxide etch (BOE) then loaded inside the MBE immediately after. Thin films of CaF$_2$ were grown epitaxially on Si(111) substrate at $600^\circ$C, with thickness ranging from $12$ to $100$ nm. \cite{caf2} After the deposition of CaF$_2$ and lowering the substrate temperature to $196^\circ$C, growth of Bi$_2$Se$_3$ was done following the two-step growth procedure on CaF$_2$ as described above. We show in Fig.~\ref{tem} a cross-sectional transmission electron microscopy (TEM) image of a $5$ QL Bi$_2$Se$_3$ deposited on CaF$_2$/Si. The $12$ nm thick CaF$_2$ buffer layer grew epitaxially on the Si$(111)$ substrate. And clear quintuple layers of Bi$_2$Se$_3$ can be seen, with the thickness of $1$ QL about $1$ nm. Bi$_2$Se$_3$ films on CaF$_2$/Si were also found to have high crystallinity and characteristic WAL behavior at low temperatures, indistinguishable from films grown on pure CaF$_2$ substrates. The magnetoresistance of a $10$ QL Bi$_2$Se$_3$ film on CaF$_2$/Si was plotted in Fig.~\ref{mr}(a), in which quantum oscillations, measured at $2$ K, were clearly visible, especially after a smooth background was subtracted, as shown in Fig.~\ref{mr}(b). From the periods of the oscillations, we obtain a Fermi wave number of $k_F = 5.04 \times 10^6$ cm$^{-1}$ and a surface carrier density of $n_s = 2.00 \times 10^{12}$ cm$^{-2}$. Important for our present discussion and relevant for the WL and WAL analysis of TI thin films, on both CaF$_2$ and CaF$_2$/Si, is that the pure WAL appears with a much lower scattering rate. Since Quantum oscillations are suppressed exponentially for higher scattering rate, we determine that the oscillations indeed come from the surface states. Moreover, Hall effect measured in our films indicates two dimensional carrier density similar to previously measured densities, \cite{QikunXue1,QikunXue2,MBE1,MBE2,MBE3,MBE4,Ando} while the carrier density obtained from the quantum oscillations are one order of magnitude smaller. We also noted that our values of $k_F$ and $n_s$ are very close to the values reported for Bi$_2$Se$_3$ films on Al$_2$O$_3$ substrates recently for similar thicknesses. \cite{Ando} %Thus we conclude that our method of growth primarily improves the quality of the surface and the surface state, which is of utmost importance for possible applications. 

In conclusion, we have demonstrated that a RF cracker cell for Se source can significantly increase the chemical reactivity of evaporated Se by dissociating Se molecules to Se atoms, which enables the deposition of Bi$_2$Se$_3$ using a Bi-to-Se ratio close to exact stoichiometry. In high quality ultrathin Bi$_2$Se$_3$ films on CaF$_2 (111)$ substrates deposited with such a RF cracker cell for atomic Se source, we observed the emergence of weak localization in addition to the commonly observed weak-antilocalization effects, as predicted by theories. After depositing a thin CaF$_2(111)$ layer epitaxially on Si$(111)$ substrate, we were able to grow high quality Bi$_2$Se$_3$ films on CaF$_2$/Si, in which quantum oscillations of the surface states were observed. Those devices can now be gated, paving the way towards the realization of a fully functioning back-gated topological insulator device with tunable Fermi level. 

This work was supported by fundings from FENA and DARPA. Merav Dolev and Alexander Palevski were partially funded by a DoE Seed funding for studying TI. We would like to acknowledge Ann Marshall for her help with TEM. We also thank Ko Munakata, Gerwin Hassink, James Analytis, Garrett Hayes, Malcolm Beasley, Jing Wang, Xiaoliang Qi, Shoucheng Zhang, Nicholas Breznay, Nai-Chang Yeh and Kang Wang for useful discussions.

\bibliographystyle{plain}

\end{document}